\documentstyle[aps,twocolumn,epsf]{revtex}
\begin{document}

\author{Itzhak Bars}
\title{   {\small \noindent hep-th/9411078 \hfill USC-94/HEP-B2 }  \bigskip\\
Folded Strings in Curved Spacetime}
\address{Department of Physics and Astronomy\\
University of Southern California\\
Los Angeles, CA 90089-0484}
\date{November 10, 1994} {
\maketitle

\begin{abstract}
\centerline{\bf Abstract}\bigskip

Two dimensional classical string theory is solved in any curved spacetime.
The complete spacetime required to describe the classical {\it string}
motions turns out to be larger than the global space required by complete
{\it particle} geodesics. The solutions are fully classified by their
behavior in the asymptotically flat region of spacetime. When the curvature
is smooth, the string solutions are deformed folded string solutions as
compared to flat spacetime folded strings that were known for 19 years.
However, surprizing new stringy behavior becomes evident at curvature
singularities such as black holes. The global properties of the classical
string theory require that the ``bare singularity region'' of the black hole
be included along with the usual black hole spacetime. The mathematical
structure needed to describe the solutions include a recursion relation that
is analogous to the transfer matrix of lattice theories. This encodes
lattice-like properties on the worldsheet on the one hand and the geometry
of spacetime on the other hand. A case is made for the presence of folded
strings in the {\it quantum} theory of non-critical strings for $d\geq 2$.
\end{abstract}



\section{Introduction}

The original physical motivations for studying string theory were: (1)
understanding unification of forces including quantum gravity, and (2)
understanding the Standard Model. In recent years it has become increasingly
evident that these goals should be re-examined in the presence of curved 4D
space-time string backgrounds. The gauge symmetries and spectrum of quark +
lepton families, which are the main ingredients of the Standard Model, were
probably fixed during the stringy early times in the evolution of the
Universe. At such times 4D space-time was curved. Since curvature
contributes to the central charge, duality and other topological aspects of
String Theory, it is likely that the predictions of String Theory in curved
4D spacetime are quite different than the flat 4D approach. Therefore,
String Theory in curved space-time must be better understood in order to
discuss its connections to low energy physics.

One should consider all kinds of curved backgrounds, not only the
traditional cosmological backgrounds, since the passage from curved
space-time to flat space-time may involve various phase transitions,
including inflation of a small region of the original curved universe to
today's large universe that is essentially homogeneous and flat. Despite the
curvature during the early universe, one can identify the states that play a
role in low energy physics purely on the basis of symmetries: The {\it gauge
bosons}, and {\it chiral families} of quarks and leptons in a small region
of the {\it 4D early curved universe} would become the ``massless''
particles observed in today's inflated 4D flat universe. Therefore, one can
search for the ``right'' curved space model on the basis of its symmetries
and ``low energy spectrum''. Such a scenario (including inflation, and
tracking the low energy states) is technically little understood although
logically and intuitively it is highly plausible. The possibility of such a
scenario suggests that curved space-time string theory deserves intensive
study.

In addition, the issues surrounding gravitational singularities should be
answered in the context of curved space-time string theory, as it is the
only known theory of quantum gravity.

With these questions in mind, we have been pursuing a program of building
and analyzing {\it exactly solvable models} of string theory in curved
spacetime. The main tool introduced in \cite{ibhetero}\cite{BN} is the G/H
gauged WZW model based on non-compact groups, such that the coset contains
{\it a single time coordinate}. The geometrical properties of such models
started to be understood when the $SL(2,R)/R$ case with $k=9/4$ \cite{BN}
was interpreted as a string moving in a black hole background \cite{wit}.
Since then a lot of progress was made on the construction of exact conformal
field theory models for bosonic, supersymmetric and heterotic strings in
curved spacetime. Some exact results on the geometrical properties of the
models (to all orders in $\alpha ^{\prime })$ were derived (\cite{ibsfglo}-%
\cite{sfrecent}). In such models various questions can be investigated. In
particular, the ``low energy spectrum'' can be computed group theoretically
by using the unitary representations of the relevant non-compact group \cite
{ibhetero}\cite{ibspectrum}.

More recently, it became apparent that the solution of the classical
equations of such models would shed more light on their physical
interpretation. This is relevant to fundamental questions of singularities
in gravitational physics, as well as stringy questions about the early
universe and its influence on the low energy spectrum of quarks and leptons.
The principles of the full classical stringy solution for any gauged WZW
model were obtained in general terms in \cite{ibsfclass}. The specialization
to particle geodesics was given explicitly in \cite{ibsfglo}. A more
detailed exploration of the general 2D classical string theory in any curved
spacetime (i.e. not only WZW models) was done in \cite{ibjs}. In 2D the only
non-trivial stringy solutions turn out to be necessarily folded strings, and
therefore folded strings are the only path toward analyzing stringy
questions in a 2D toy model. The analysis of \cite{ibjs} showed that,
surprizingly, strings behave quite differently than particles in the
vicinity of singularities such as a black hole or a big bang.

In the present paper we will discuss more generally the classical solutions
of 2D string theory in any curved spacetime. We will systematize the results
of \cite{ibjs} by providing closed compact formulas for the classical
solutions in any 2D spacetime metric $G_{\mu \nu }(x(\tau ,\sigma )).$ The
method of solution introduces a lattice on the worldsheet (not on spacetime)
as a mathematical convenience, and then defines a sort of ``transfer
matrix'' that incorporates properties of the worldsheet lattice on the one
hand and of the curved target spacetime geometry on the other hand. This
``transfer matrix'' is a recursion relation that builds a single continuous
string solution by ``weaving'' together the string solutions in the patches
of the worldsheet. A very important property of the transfer matrix turns
out to be certain minimal area {\it conservation laws} that must be obeyed
by the string motion. The general formulas are then applied to the 2D black
hole and the 2D cosmological de Sitter cases. Due to the conservation laws,
it turns out that, in contrast to particle geodesics, the string geodesics
require the inclusion of the ``bare singularity region'' of the black hole
for a complete spacetime. A string falling into a black hole tunnels into
the bare singularity region, and its motion cannot be described{\it \
classically }without the inclusion of this{\it \ ``dual region''. }%
Therefore, the usual description of the black hole spacetime is simply
incomplete even in classical physics. This may imply that additional new
phenomena must be taken into account at any black hole for the correct
description of the physics.

In addition to the interest in singular gravitational behavior there has
also been a long-standing interest in exploring consistent generalizations
of non-critical strings with the hope that they may be relevant for some
branch of physics. Folded strings fall into this category, especially in the
area of string-QCD relations. Therefore our work touches on two aspects of
string theory: (i) strings in curved space-time and (ii) folded strings.

The main discussion of the classical folded string solutions in 2D is given
in sections 2-6. In sections 7,8 a case is made for the presence of
folded strings in the quantum theory of non-critical strings and in higher
dimensions, by clarifying the reasons for their absence in the standard
methods of conformal field theory and other approaches. Comments and
conclusions appear in section 9.

\section{Classical solution}

In \cite{ibjs} the complete set of solutions of two dimensional classical
string theory were constructed for {\it any 2D curved spacetime}. The
classical action is given by $\smallint d^2\sigma \,G_{\mu \nu }(x)\partial
_{+}x^\mu \partial _{-}x^\nu .$ In 2D $B_{\mu \nu }(x)$ can be eliminated
since it produces a total derivative in the action, and in the classical
theory the dilaton is absent. The most general metric can always be
transformed into the conformal form $G_{\mu \nu }=\eta _{\mu \nu }G(x).$
Then the most general 2D {\it classical} string equations of motion $%
\partial _{+}\partial _{-}x^\mu +\Gamma _{\nu \lambda }^\mu \partial
_{+}x^\nu \partial _{-}x^\lambda =0,$ and conformal (Virasoro) constraints $%
\partial _{\pm }x^\mu \partial _{\pm }x^\nu G_{\mu \nu }(x)=0$ (vanishing
stress tensor) take the form
\begin{equation}
\begin{array}{c}
\partial _{+}(G\,\,\partial _{-}u)+\partial _{-}(G\,\,\partial _{+}u)=\frac{%
\partial G}{\partial v}(\partial _{+}u\partial _{-}v+\partial _{+}v\partial
_{-}u) \\
\partial _{+}(G\,\,\partial _{-}v)+\partial _{-}(G\,\,\partial _{+}v)=\frac{%
\partial G}{\partial u}(\partial _{+}u\partial _{-}v+\partial _{+}v\partial
_{-}u) \\
\partial _{+}u\partial _{+}v=0=\partial _{-}u\partial _{-}v\,\,\,,
\end{array}
\label{stringeqs}
\end{equation}
where we have used the target space lightcone coordinates $\,\,u(\sigma
^{+},\sigma ^{-})=\frac 1{\sqrt{2}}(x^0+x^1),\,\,v(\sigma ^{+},\sigma ^{-})=%
\frac 1{\sqrt{2}}(x^0-x^1),$ and the world sheet lightcone coordinates $%
\sigma ^{\pm }=(\tau \pm \sigma )/\sqrt{2},\,\,\,\partial _{\pm }=(\partial
_\tau \pm \partial _\sigma )/\sqrt{2}.$

Since a typical string state is massive, one should expect that the string
will follow {\it on the average} the trajectory of a {\it massive} particle.
Therefore, to understand the average behavior of the string geodesic it is
useful to first consider the solution for the geodesic of a massive
particle. The particle geodesic equations follow from the above ones by
dimensional reduction. That is, by dropping the $\sigma $ dependence, $%
\partial _{\pm }\rightarrow \partial _\tau ,$ these equations reduce to the
point particle geodesic equations. For particles, the last line in (\ref
{stringeqs}) imposes the condition for a null geodesic, which is too
restrictive for our purpose. If this condition is modified to
\begin{equation}
G\dot u\dot v=\frac{m^2}2
\end{equation}
then (\ref{stringeqs}) become the equations for a timelike geodesic for a
massive particle with mass $m.$ The zero mass limit may also be considered
at the end. We will provide the explicit solutions to the particle as well
as the string equations. As discovered in \cite{ibjs} there are additional
{\it stringy phenomena due to the wave nature} that cannot be seen in the
particle solution, and therefore it is useful to contrast the string
solutions with the particle solutions.

In flat space-time the solutions are given in terms of arbitrary left-moving
and right-moving functions $x_L^\mu (\sigma ^{+}),\,x_R^\mu (\sigma ^{-})$
\begin{equation}
x^\mu (\tau ,\sigma )=x_L^\mu (\sigma ^{+})+x_R^\mu (\sigma ^{-}).
\end{equation}
As shown by BBHP in several gauges \cite{bbhp1}\cite{bbhp2}, the constraints
are also satisfied provided\footnote{%
Although the original BBHP solutions were for open strings, the same
solutions also apply to closed strings by simply taking independent
functions $f,g$ for left movers and right movers.}
\begin{equation}
\begin{array}{c}
u=u_0+\frac{p^{+}}2\left[ (\sigma ^{+}+f(\sigma ^{+}))+(\sigma ^{-}-g(\sigma
^{-}))\right] \\
v=v_0+\frac{p^{-}}2\left[ (\sigma ^{+}-f(\sigma ^{+}))+(\sigma ^{-}+g(\sigma
^{-}))\right]
\end{array}
\label{general}
\end{equation}
where $f(\sigma ^{+})$ and $g(\sigma ^{-})$ are any two {\it periodic
functions, } $f(\sigma ^{+})=f(\sigma ^{+}+\sqrt{2})$, $g(\sigma
^{-})=g(\sigma ^{-}+\sqrt{2}),$ with slopes $f^{\prime }(\sigma ^{+})=\pm 1$
and $g^{\prime }(\sigma ^{-})=\pm 1.$ The slopes can change discontinuously
any number of times at arbitrary locations $\sigma _i^{+},\sigma _j^{-}$
within the basic intervals $-1/\sqrt{2}\leq \sigma ^{\pm }\leq 1/\sqrt{2}$
(and then repeated periodically), but the functions $f,g$ are continuous at
these points. The discontinuities in the slopes are allowed since the
equations of motion are first order in either $\partial _{+}$ or $\partial
_{-}$. The number of times the slope changes in the basic interval
corresponds to the number of folds for left movers and right movers
respectively. The simplest BBHP\ solution is the so called yo-yo solution
given by $f=|\sigma ^{+}|_{per}$ and $g=|\sigma ^{-}|_{per}$ which are the
periodically repeated absolute value. These solutions describe folded
strings, with the folds oscillating against each other, and moving at the
speed of light. Examples are plotted in figures 1,2.


In Fig.1 one sees the yo-yo solution with equal periods for $|\sigma
^{+}|_{per}$, and $|\sigma ^{-}|_{per}$ . Fig. 2 is generated by taking the
period of $|\sigma ^{-}|_{per}$ to be half of that of $|\sigma ^{+}|_{per}$.
In Fig.2 the string has two folds plus an additional critical point moving
at the speed of light that becomes a fold for part of the motion.


As discovered in {\cite{ibjs} }the {\it complete} set of classical solutions
in curved spacetime are classified by their behavior in the asymptotically
flat region of spacetime $G(u,v)\rightarrow 1,$ where they tend to the {\it %
complete} set folded string solutions of BBHP\ given in (\ref{general}) as
boundary conditions\footnote{%
A coordinate transformation is sometimes needed to obtain the standard flat
metric $G=1$ in some flat regions. For example for the 2D black hole metric $%
ds^2=dudv\,(1-uv)^{-1}$ there are two flat regions. The first is near the
horizon $u=0$ or $v=0,$ the second is far away $u\rightarrow \infty $ or $%
v\rightarrow \infty .$ For the second one, we must make a coordinate
transformation to obtain the standard flat metric $ds^2=\frac{du}u\frac{dv}v%
=dUdV,$ with $U=\ln u,\,V=\ln v.$ The coordinate transformation must first
be applied before connecting to the BBHP solutions as boundary conditions.}.
The curved space solutions are given in the form of a map from the world
sheet to target spacetime. As a mathematical convenience the world sheet is
divided into lattice-like patches, with a map associated with each patch.
The world-sheet lattice structure is determined by the sign patterns of ($%
f^{\prime },g^{\prime })=(\pm ,\pm )$ inherent in the BBHP solutions, thus
the lattice is dictated by the boundary conditions in the asymptotically
flat region of spacetime $G(u,v)\rightarrow 1$. We emphasize that the
lattice is on the world-sheet, not in curved spacetime, and it is only a
mathematical tool to keep track of patches. In each patch of the lattice one
set of signs holds, hence there are 4 types of patches called $A,B,C,D.$ For
each such patch there is a solution of the equations of motion that is valid
within the patch. The forms of the solutions in a group of neigboring $%
A,B,C,D$ patches labelled by an integer $k$ are (see eq.(\ref{soll}) for an
example of a pattern of patches)
\begin{equation}
\begin{array}{l}
A:\quad u=U_k(\sigma ^{+}),\quad v=U_k(\sigma ^{-}) \\
B:\quad u=U_k(\sigma ^{-}),\quad v=V_k(\sigma ^{+}) \\
C:\quad u=u_k,\quad v=W[\alpha _k(\sigma ^{+})+\beta _k(\sigma ^{-}),u_k] \\
D:\quad u=\bar W[\alpha _k(\sigma ^{-})+\beta _k(\sigma ^{+}),v_k],\quad
v=v_k\,\,\,,
\end{array}
\label{fourr}
\end{equation}
where the functions $U_k(\sigma ^{\pm }),V_k(\sigma ^{\pm })$, $\alpha
_k(\sigma ^{\pm })$, $\beta _k(\sigma ^{\pm })$ and the constants $u_k,v_k$
are given by a recursion relation whose form depends on the metric $G.$ It
is easy to verify that, independently of the recursion relation, the forms
listed in (\ref{fourr}) solve the differential equations for any $U_k(\sigma
^{\pm }),V_k(\sigma ^{\pm })$, $\alpha _k(\sigma ^{\pm })$, $\beta _k(\sigma
^{\pm })$ provided the functions $W_k(\sigma ^{+},\sigma ^{-}),\bar W%
_k(\sigma ^{+},\sigma ^{-})$ are defined by inverting the following
relations
\begin{eqnarray}
F(u_k,W_k) &\equiv &\int^{W_k}dv^{\prime }G(u_k,v^{\prime })=\alpha
_k(\sigma ^{+})+\bar \beta _k(\sigma ^{-}), \\
\bar F(\bar W_k,v_k) &\equiv &\int^{\bar W_k}du^{\prime }G(u^{\prime },v_k)=%
\bar \alpha _k(\sigma ^{-})+\beta _k(\sigma ^{+}).
\end{eqnarray}
This is the complete set of solutions\footnote{%
This set of solutions were noticed independently in \cite{ibsfclass}\cite
{ibberkeley} and \cite{deveg}, but the authors of \cite{deveg} thought that
there are no stringy solutions. They did not realize that the validity of
these solutions is limited to patches of the worldsheet, and they assumed
that the stringy solutions discussed in \cite{ibjs} and here are gauged away
by using the remaining conformal invariance.}. So, for a given metric $%
G(u,v) $ there exists the functions $F(u,v)$ and $\bar F(u,v)$ such that
their partial derivatives reproduce the metric
\begin{equation}
\frac{\partial F(u,v)}{\partial v}=G(u,v)=\frac{\partial \bar F(u,v)}{%
\partial u}.  \label{FGF}
\end{equation}
and for each metric $G$ we have the relations
\begin{equation}
\begin{array}{c}
F(u_0,v)=\alpha +\bar \beta \quad \leftrightarrow \quad v=W(\alpha +\bar
\beta ,u_0), \\
\bar F(u,v_0)=\bar \alpha +\beta \quad \leftrightarrow \quad u=\bar W(\bar
\alpha +\beta ,v_0),
\end{array}
\label{FW}
\end{equation}
that help define the solutions $C,D$ in terms of the arbitrary functions $%
\alpha (\sigma ^{+})$, $\,\beta (\sigma ^{+})$, $\,\bar \alpha (\sigma
^{-}),\,\bar \beta (\sigma ^{-}).$ Consider the following three cases as
illustrations

\begin{enumerate}
\item  Flat metric $ds^2=du\,dv$ :
\begin{equation}
\begin{array}{l}
F=u_0+v=\alpha +\bar \beta , \\
\quad quad \leftrightarrow \quad v=W=\alpha +\bar \beta -u_0, \\
\bar F=u+v_0=\bar \alpha +\beta , \\
\quad \quad \leftrightarrow \quad u=\bar W=\bar \alpha +\beta -v_0.
\end{array}
\label{FWflat}
\end{equation}

\item  SL(2,R)/R black hole metric $ds^2=(1-uv)^{-1}du\,dv$:
\begin{equation}
\begin{array}{l}
F=-u_0^{-1}\ln (1-u_0v)=\alpha +\bar \beta \quad  \\
\quad \leftrightarrow \quad v=W=u_0^{-1}\{1-\exp [-u_0(\alpha +\bar \beta
)]\}, \\
\bar F=-v_0^{-1}\ln (1-uv_0)=\bar \alpha +\beta \quad  \\
\quad \leftrightarrow \quad u=\bar W=v_0^{-1}\{1-\exp [-v_0(\bar \alpha
+\beta )]\}.
\end{array}
\label{FWsl}
\end{equation}

\item  Cosmological (de Sitter) metric $ds^2=dt^2-e^{2Ht}dr^2=\frac 4{H^2}%
(u+v)^{-2}du\,dv$:
\begin{equation}
\begin{array}{l}
F=-(u_0+v)^{-1}=\alpha +\bar \beta  \\
\quad \quad \leftrightarrow \quad v=W=-(\alpha +\bar \beta )^{-1}-u_0, \\
\bar F=-(u+v_0)^{-1}=\bar \alpha +\beta  \\
\quad \quad \leftrightarrow \quad u=\bar W=-(\bar \alpha +\beta )^{-1}-v_0.
\end{array}
\label{FWdesitter}
\end{equation}
\end{enumerate}

The reader can verify that the flat spacetime BBHP solutions given in (\ref
{general}) take the 4 forms of (\ref{fourr}) in regions of $\sigma ^{\pm }$
where the 4 types of sign patterns ($f^{\prime },g^{\prime })=(\pm 1,\pm 1)$
hold. The form (\ref{FWflat}) looks more general than the BBHP solution
because there still are boundary conditions and gauge freedom that will be
fixed later. Then it agrees precisely with (\ref{general}) as seen below.
Intuitively we expect that in the presence of small curvature the physical
character of the solution is similar to the folded string motions
illustrated in Figs.1,2. As the curvature increases smoothly, except for the
deformations due to curvature, the motions must also be similar. The
question is what happens when there are curvature singularities?

At the boundaries of each patch continuity conditions must be imposed. This
produces a recursion relation that describes the motion of the string as the
proper time $\tau $ increases (see below). The recursion relation, which is
analogous to a ``transfer matrix'' of lattice theories, connects the maps in
different patches into a single continuous map from the worldsheet to
spacetime. Thus, the functions $U_k(\sigma ^{\pm }),V_k(\sigma ^{\pm })$, $%
\alpha _k(\sigma ^{\pm })$, $\beta _k(\sigma ^{\pm })$ in the various
patches get related to each other. This ``transfer matrix'' encodes the
properties of the world sheet lattice on the one hand and the geometry of
spacetime on the other hand. {Thus, }lattices on the world-sheet plus
geometry in space-time lead to ``transfer matrices''. This seems to be a
rich area of mathematical physics to explore in more detail in the future.
Here we will derive the transfer matrix for one such lattice.

Recall that the lattice is dictated by the nature of the solution (\ref
{general}) in the asymptotically flat region of target spacetime. As an
example we consider the simplest yo-yo solution as a boundary condition.
This defines the sign patterns according to the slopes of the periodic
functions $|\sigma ^{+}|_{per}$ and $|\sigma ^{-}|_{per},$ and the following
lattice emerges from the periodic behavior of these functions . The world{\
\ sheet is labelled by $\sigma $ horizontally and by $\tau $ vertically.
Increasing values of }$k$ correspond to increasing values of $\tau .$ {%
Periodicity in }$\sigma $ is imposed, hence the world sheet is a cylinder. {%
It is sliced by equally spaced $45^o$ lines that form a light-cone lattice
in $\sigma ^{\pm }$. The crosses in the diagram represent the corners of the
cells on the world sheet.} {\tiny
\begin{equation}
\begin{array}{ccccc}
\begin{array}{c}
\sigma =0 \\
\vdots
\end{array}
&
\begin{array}{c}
\sigma =1 \\
\vdots
\end{array}
&
\begin{array}{c}
\sigma =2 \\
\vdots
\end{array}
&
\begin{array}{c}
\sigma =3 \\
\vdots
\end{array}
&
{^{\sigma =4=0} \atopwithdelims.. \vdots }
\\
\times &
\begin{array}{c}
\begin{array}{c}
U_{k+2}(\sigma ^{+}) \\
V_{k+2}(\sigma ^{-})
\end{array}
\end{array}
& \times &
\begin{array}{c}
U_{k+2}(\sigma ^{-}) \\
V_{k+2}(\sigma ^{+})
\end{array}
& \times \\
&  &  &  &  \\
\cdots
\begin{array}{c}
u_{k+1} \\
W_{k+1}(\sigma ^{+},\sigma ^{-})
\end{array}
& \times &
\begin{array}{c}
\bar W_{k+1}(\sigma ^{+},\sigma ^{-}) \\
v_{k+1}
\end{array}
& \times & \cdots \\
&  &  &  &  \\
\times &
\begin{array}{c}
U_{k+1}(\sigma ^{-}) \\
V_{k+1}(\sigma ^{+})
\end{array}
& \times &
\begin{array}{c}
U_{k+1}(\sigma ^{+}) \\
V_{k+1}(\sigma ^{-})
\end{array}
& \times \\
&  &  &  &  \\
\cdots
\begin{array}{c}
\bar W_k(\sigma ^{+},\sigma ^{-}) \\
v_k
\end{array}
& \times &
\begin{array}{c}
u_k \\
W_k(\sigma ^{+},\sigma ^{-})
\end{array}
& \times & \cdots \\
&  &  &  &  \\
\times &
\begin{array}{c}
U_k(\sigma ^{+}) \\
V_k(\sigma ^{-})
\end{array}
& \times &
\begin{array}{c}
U_k(\sigma ^{-}) \\
V_k(\sigma ^{+})
\end{array}
& \times \\
&  &  &  &  \\
\cdots
\begin{array}{c}
u_{k-1} \\
W_{k-1}(\sigma ^{+},\sigma ^{-})
\end{array}
& \times &
\begin{array}{c}
\bar W_{k-1}(\sigma ^{+},\sigma ^{-}) \\
v_{k-1}
\end{array}
& \times & \cdots \\
\vdots & \vdots & \vdots & \vdots & \vdots
\end{array}
\label{soll}
\end{equation}
}

The transfer matrix for this ``yo-yo lattice'' was discussed in {{\cite{ibjs}
for the }}$SL(2,R)/R$ metric$.$ Here we first give a compact general formula
for any metric $G(u,v)$ and then specialize it to the examples of flat
spacetime, the $SL(2,R)/R$ {black hole space-time, {and the cosmological de
Sitter space-time}}.

\section{General spacetime}

{\ {\ The continuity at the corners that join the $A,B$ cells is
automatically insured by the use of the same functions $U_k(z),V_k(z)$ to
describe the $A,B$ solutions, but with different arguments $z=\sigma ^{\pm }$
that alternate between neighboring cells\footnote{%
It is also possible to take different functions in the $A,B$ patches and
only require continuity at the boundaries. However, this freedom has no
physical meaning since it can be changed by a conformal gauge
transformation. Indeed, we have already used part of the remaining conformal
invariance in choosing the forms in (\ref{fourr}) to make the functions in
neighboring $A,B$ patches the same. The same reasoning apllies also to the $%
C,D$ patches.}. Continuity at the boundaries between $A,B$ cells and $C,D$
cells requires
\begin{equation}
\begin{array}{c}
U_{k+1}(-1/\sqrt{2})=U_k(1/\sqrt{2})=u_k, \\
V_{k+1}(-1/\sqrt{2})=V_k(1/\sqrt{2})=v_k.
\end{array}
\label{boundary}
\end{equation}
where the $(u_k,v_k)$ are constants. Similarly, by taking into account the
relations (\ref{FW}) at these bo{und}aries one can construct the functions $%
W_k(\sigma ^{+},\sigma ^{-}),\bar W_k${$(\sigma ^{+},\sigma ^{-})$} for the $%
C,D$ cells in terms of the functions {$U_k(\sigma ^{\pm }),V_k$}}}$(\sigma
^{\pm })$
\begin{equation}
\begin{array}{l}
^{W_k=W\left[ \left( F(u_k,V_k(\sigma ^{+}))+F(u_k,V_k(\sigma
^{-}))-F(u_k,v_{k-1})\right) ,u_k\right] }\\
_{\bar W_k=\bar W\left[ \left( \bar F(U_k(\sigma ^{+}),v_k)+\bar F(U_k(\sigma
^{-}),v_k)-\bar F(u_{k-1},v_k)\right) ,v_k\right] .}
\end{array}
\label{wwbar}
\end{equation}

{{Evaluating these at the lower (i.e. past) boundaries of the $C,D$ cells,
using $V_k(-1/\sqrt{2})=v_{k-1},$ $U_k(-1/\sqrt{2})=u_{k-1}$ , the boundary
matching is insured by the fact that $F$ and $W$ are inverses of each other
(see eq.(\ref{FW}-\ref{FWdesitter})
\begin{equation}
W\left( F(u_k,V_k(z)),u_k\right) =V_k(z),  \label{lower}
\end{equation}
and similarly for $U_k(z)$. At the upper (i.e. future) boundaries of the $%
C,D $ cells the boundary matching gives a {\it recursion relation}}}

{{{{{\
\begin{equation}
\begin{array}{l}
^{V_{k+1}(z)=W\left[ \left( F(u_k,V_k(z))+F(u_k,v_k)-F(u_k,v_{k-1})\right)
,u_k\right]} \\  _{U_{k+1}(z)=\bar W\left[ \left( \bar F(U_k(z),v_k)+\bar
F(u_k,v_k)-\bar F(u_{k-1},v_k)\right) ,v_k\right] .}
\end{array}
\label{recursion}
\end{equation}
}}} where $z=\sigma ^{\pm }.\,$This recursion may be viewed as a {\it %
transfer} \thinspace operation in proper time $\tau \rightarrow \tau +2$,
for any $\sigma $, and is quite analogous to the concept of the ``transfer
matrix'' in lattice theories. The recursion leads to the solution of all the
$U_k(\sigma ^{\pm }),\,V_k(\sigma ^{\pm })$ in terms of $U_0(z),\,V_0(z)$,
that describe initial conditions at $\tau =0$. } }

{\ {\ By evaluating the recursion relation (\ref{recursion}) at the
boundaries of each cell $z=\pm 1/\sqrt{2}$ and using the values (\ref
{boundary}) at the boundaries, one finds a recursion relation for the
constants $(u_k,v_k)$%
\begin{equation}
\begin{array}{l}
v_{k+1}=W\left[ \left( 2F(u_k,v_k)-F(u_k,v_{k-1})\right) \,,u_k\right] , \\
u_{k+1}=\bar W\left[ \left( 2\bar F(u_k,v_k)-\bar F(u_{k-1},v_k)\right)
\,,v_k\right] .
\end{array}
\label{constants}
\end{equation}
The solution of this recursion relation requires $4$ initial constants $%
u_0,v_0$, $u_{-1}$, $v_{-1}$
\begin{equation}
\begin{array}{c}
U_0(-1/\sqrt{2})=u_{-1}\quad U_0(1/\sqrt{2})=u_0\quad , \\
V_0(-1/\sqrt{2})=v_{-1}\quad V_0(1/\sqrt{2})=v_0\quad .
\end{array}
\label{init}
\end{equation}
Therefore, the positions $(u_k,v_k)$ are fully determined in curved
space-time in terms of 4 initial constants. The counting of initial
parameters is right: since there are just two dynamical folds, specifying
their initial positions and velocities corresponds to just 4 parameters. } }

{\ {\ The constants $(u_k,v_k)$ are sufficient to describe the physical
motion of the folds (or end points), as well as the whole string, as
follows. Consider the diagram of eq.(\ref{soll}). At any $\tau $ the
trajectories of the folds are parametrized by the vertical lines that pass
through $\sigma =0,2$ on the world sheet (and their periodic repetitions at\
$\sigma =4l,\,4l+2)$. Likewise, vertical lines that pass through the crosses
located at $\sigma =1,3$ (and their periodic repetitions at $\sigma
=4l+1,\,4l+3)$ parametrize the trajectory of the midpoint between the folds.
The center of mass of the string coincides with these midpoints. As $\tau $
increases one can read off the space-time trajectories of the center of mass
and of the folds by moving upward along the vertical lines in the diagram.
For example, consider the $\sigma =0$ fold: during 2$k-2\leq \tau \leq 2k$
it remains at constant $u=u_{k-1}$ while the value of $v=W_{k-1}$ increases
from $v=v_{k-2}$ to $v=v_k$. Between 2$k\leq \tau \leq 2k+2$ it remains at
constant $v=v_k$ while the value of $u=\bar W_k$ increases from $u=u_{k-1}$
to $u=u_{k+1},$ etc. In a similar way the trajectory of the second fold and
of the center of mass are read off directly from the diagram in eq.(\ref
{soll}). The space-time trajectories of these points are plotted in a $(u,v)$
plot in Fig.3.}}


{{\ } \ {\ The detailed motion of the intermediate points of the string at
any $\sigma $ are described by the functions $U_k,V_k,W_k,\bar W_k$ as
indicated on the diagram (\ref{soll}) and mapped on Fig.3. The space-time
trajectories of folds or end points that are the images of $\sigma =0,2$ are
physical and cannot depend on conformal reparametrizations. Indeed, as seen
from the above solution there is no freedom in the choice of the constants $%
(u_k,v_k)$ except for the initial values (\ref{init}). On the other hand,
the motion of the rest of the string is gauge dependent at intermediate
points $\sigma $ (because of reparametrizations), and therefore it depends
on the choice of $U_0(z),V_0(z)$ that have remained unspecified. However,
once the motion of the end points is plotted, it is clear from Fig.3 that
the shape of the minimal surface is already determined without needing the
details of the gauge dependent motion of the intermediate points. } }

{\ {\ The remaining conformal invariance may be used to fix the form of
these functions in the initial cell (although this is not necessary). For
the yo-yo solution the initial functions $U_0(z),\,V_0(z)$ need not contain
more than $4$ constants that are related to the initial positions and
velocities of the two folds. Therefore, the simplest gauge fixed form is
\begin{equation}  \label{initial}
\begin{array}{c}
U_0(z)=\frac 12(u_0+u_{-1})+\frac 1{\sqrt{2}}(u_0-u_{-1})\,z_{per} \\
V_0(z)=\frac 12(v_0+v_{-1})+\frac 1{\sqrt{2}}(v_0-v_{-1})\,z_{per},
\end{array}
\end{equation}
where $z_{per}$ is the linear function $z_{per}=z$ in the interval $-1/\sqrt{%
2}\leq z\leq 1/\sqrt{2}$, and then repeated periodically. However, any other
periodic function with the same 4 boundary constants will produce the same
physical motion for the folds. } }

{\ {\ The recursion (\ref{constants}) is the fundamental physical relation
that fully determines the motion of the yo-yo string in curved space-time.
We called it the ``transfer matrix'' in the example of the black hole worked
out in ref.\cite{ibjs}. It was found that it has certain invariances that
are valid everywhere in target space-time, including near singularities. The
invariance is related to a lattice version of the fundamental action $%
A=\smallint d^2\sigma \,G_{\mu \nu }\partial _{+}x^\mu \partial _{-}x^\nu $
that represents the minimal surface swept by the string. The lattice version
of the minimal surface is expressed in terms of the constants $(u_k,v_k)$ ,
and its value for one period turns out to be a constant of motion. Explicit
expressions for this lattice action will be given for specific metrics in
the following sections. For every metric $G$ one can find a lattice version
of the action $A$ that is an invariant under the recursion (\ref{constants}%
). The invariance is valid even in the vicinity of singularities in
space-time ( i.e. when $G(u_k,v_k)$ grows) and helps in the understanding of
new stringy phenomena. For example, it was found that classical strings can
tunnel to regions of space-time (such as the bare singularity region of a
black hole) that are forbidden to particle geodesics. Such a surprising
motion of a string may be thought of as the analog of the diffraction of
light around corners, that is possible for classical waves, but is
impossible for particle trajectories. } }

{\ {\ In this section we constructed the yo-yo solution in any curved
space-time given by $G$. In a similar way one may consider more complicated
solutions with many folds. The general boundary condition near $G\rightarrow
1$ given by (\ref{general}), with any number folds, defines a pattern of $%
A,B,C,D$ on the world sheet that corresponds to the regions of $(\sigma
^{+},\sigma ^{-})$ that have definite signs of $f^{\prime },g^{\prime }$ for
some choice of $f,g.$ The pattern must be periodic horizontally, with a
period of $\sigma \rightarrow \sigma +4$, to insure periodicity. This
generalizes the lattice in the diagram of (\ref{soll}). By virtue of the
BBHP construction, any of these generalized patterns is guaranteed to
correspond to strings that propagate forward in time. Then there remains to
carry out the matching of the functions at the boundaries. This would give
generalizations of the recursion relations and transfer matrices discussed
above. It seems that this is a very rich area for mathematical physics,
since one may explore relations between geometries defined by metrics $G$,
lattices, and transfer matrices. It is clear that the general behavior of
the minimal surface that emerges from this procedure has to be quite similar
to the one in flat space-time (which is already given by the choice of $f,g$%
), except for the deformations due to curvature and singularities. Moreover,
it seems that the main physical stringy features related to the curvature
and/or singularity structure of space-time may already be extracted from the
yo-yo solution that has only two folds. } }

{\ {\ We now apply the general yo-yo results to several specific metrics and
construct explicitly the corresponding ``transfer matrices'', their
invariants, and the corresponding string solutions. } }

\section{Flat Space-time}

{\ {\ The functions $\bar F,\bar W$ corresponding to the flat space-time
metric $G=1$ are given in (\ref{FWflat}). Using them in the general formulas
(\ref{boundary}-\ref{init}) we obtain the explicit recursion relations
\begin{equation}
\begin{array}{c}
\bar W_k=U_k(\sigma ^{+})+U_k(\sigma ^{-})-u_{k-1}, \\
U_{k+1}(z)=U_k(z)+u_k-u_{k-1} \\
u_{k+1}=2u_k-u_{k-1}
\end{array}
\label{recflat}
\end{equation}
They are solved by
\begin{equation}
\begin{array}{c}
u_k=u_0+k(u_0-u_{-1}) \\
U_k(z)=U_0(z)+k(u_0-u_{-1}) \\
\bar W_k=U_0(\sigma ^{+})+U_0(\sigma ^{-})+(k+1)(u_0-u_{-1})-u_0
\end{array}
\label{solflat}
\end{equation}
where $U_0(-1/\sqrt{2})=u_{-1},\,\,U_0(1/\sqrt{2})=u_0$, and the function $%
U_0(z)$ is arbitrary. The solutions for $V_k(z),W_k,v_k$ are obtained from
the above by replacing $U\rightarrow V$ and $u\rightarrow v$. If $%
U_0(z),V_0(z)$ are gauge fixed as in (\ref{initial}), then this solution
takes the convenient form of the BBHP yo-yo string in (\ref{general}) with }}%
$f=|\sigma ^{+}|_{per},\,\,g=|\sigma ^{-}|_{per}${{. The present form is a
generalization that permits other gauge choices. The motion of the end
points, as plotted in Fig.1 is gauge independent, but the motion of the
interior points of the string depends on the gauge choice, as expected. } }

{\ {\ Define a lattice version of the surface element $dA=d^2\sigma
\,(\partial _{+}u\,\partial _{-}v+\partial _{-}u\,\partial _{+}v)$ swept by
the string during $2k\leq \tau \leq 2k+2.$ The area of one rectangle in
Fig.1 is
\begin{equation}
dA_k=(u_k-u_{k-1})\,(v_k-v_{k-1}).  \label{minimflat}
\end{equation}
{}From the world sheet point of view this covers the image of one $A$ or $B$
cell. }}T{{he image of a $C$ or $D$ cell has zero area in target space-time
since they are mapped to the edges of the rectangle (see footnote 6 for a
related point). Consider the transformation (\ref{recflat}) as a transfer
matrix that takes the system forward in time. Under this transformation $dA_k
$ is an invariant since $dA_{k+1}=dA_k$ . This is seen by rewriting (\ref
{recflat}) in the form $U_{k+1}(z)-u_k=U_k(z)-u_{k-1}$, etc.. Therefore, we
may say that the ``transfer matrix'' for flat space-time given by (\ref
{recflat}) leaves invariant the ``lattice action density'' given by (\ref
{minimflat}). This concept generalizes to curved space-time, as seen below. }
}

\section{Black hole space-time}

{\ {\ The case of the SL(2,R)/R two dimensional black hole metric $%
ds^2=(1-uv)^{-1}du\,dv$ was already discussed in \cite{ibjs}, but here we
will show how the results of \cite{ibjs} follow from the general formulas,
and also give the additional recursion relations for $U_k,V_k,W_k,\bar W_k$
at general $k$ and general gauge that were not provided in \cite{ibjs}. } }

{\ {\ The solution for the geodesic of a massive particle was given in our
previous work \cite{ibsfglo} \cite{ibjs}. Here we rewrite it in a more
convenient form
\begin{equation}
\begin{array}{l}
u=e^{\tau \sqrt{\gamma ^2+\frac{m^2}2}}\left[
\begin{array}{l}
u_0\cosh (\gamma \tau )- \\
\frac{\left( u_0\sqrt{\gamma ^2+\frac{m^2}2}-\dot u_0\right) \sinh (\gamma
\tau )}\gamma
\end{array}
\right]  \\
{{v=e^{-\tau \sqrt{\gamma ^2+\frac{m^2}2}}\left[
\begin{array}{l}
v_0\cosh (\gamma \tau )+ \\
\frac{\left( v_0\sqrt{\gamma ^2+\frac{m^2}2}+\dot v_0\right) \sinh (\gamma
\tau )}\gamma
\end{array}
\right] ,}}
\end{array}
\label{geosl}
\end{equation}
where $u_0,v_0,\dot u_0,\dot v_0$ are initial velocities and momenta, $m$ is
the mass of the particle, and $\gamma $ is a convenient parameter
\begin{equation}
\begin{array}{l}
\gamma =\frac{\sqrt{(u_0\dot v_0+\dot u_0v_0)^2-4\dot u_0\dot v_0}}{%
2(1-u_0v_0)}, \\
{{\frac{m^2}2=\frac{\dot u_0\dot v_0}{(1-u_0v_0)}.}}
\end{array}
\label{geoslpar}
\end{equation}
In the zero mass limit either $\dot u_0=0$ or $\dot v_0=0$, and then the
solution reduces to a light-like geodesic for which either $u$ or $v$ remain
constant respectively at all times. } }

{\ {\ The singularity is at $u(\tau )v(\tau )-1=0.$ To see when the particle
hits the singularity we compute this quantity
\begin{equation}
\frac{uv-1}{u_0v_0-1}=\left[ \cosh \gamma \tau +\frac{\left( u_0\dot v_0+%
\dot u_0v_0\right) \sinh \gamma \tau }{\sqrt{(u_0\dot v_0+\dot u_0v_0)^2-4%
\dot u_0\dot v_0}}\right] ^2  \label{hit}
\end{equation}
In the massless limit this expression becomes
\begin{equation}
\frac{u(\tau )v(\tau )-1}{u_0v_0-1}=\exp \left( (u_0\dot v_0+\dot u%
_0v_0)\tau \right) ,\quad \dot u_0\dot v_0=0.  \label{hitt}
\end{equation}
It is evident that the sign of $uv-1$ cannot change as $\tau $ changes,
therefore the particle must remain in either the black hole region $uv<1$
(can cross the horizon at $u=0$ or $v=0$), or in the bare singularity region
$uv>1.$ The boundary $uv=1$ acts like an impenetrable wall from either side.
This last feature is different for the string solution. In contrast to the
point particle, the string will tunnel through the wall !! This surprising
effect was discovered in \cite{ibjs}. } }

{\ {\ It was evident from the work of \cite{ibjs} that, except for the
tunneling type phenomena, the string follows more or less the geodesic of
the {\it massive} particle. Therefore, it is useful to clarify the
properties of the geodesics of the point particle, because they depend on
the initial particle location as well as its velocity. } }

{{If the particle starts out in the ``bare singularity'' region , $u_0v_0>1$
(future or past regions)$,$ the mass formula in (\ref{geoslpar}) requires $%
\dot u_0\dot v_0<0$ and $\gamma $ is real$.$ Then the motion is governed by
hyperbolic functions, and (\ref{hit}) never vanishes. Therefore, a {\it %
massive} {\it particle, or the string, cannot hit the singularity}. In the
massless limit, according to (\ref{hitt}), the light-like geodesic will hit
the bare singularity only if it starts out with initial conditions that give
$u_0\dot v_0+\dot u_0v_0<0,$ but in any case it reaches the singularity only
at infinite proper time $\tau =\infty .$ Therefore, the ``bare singularity''
region of the SL(2,R)/R black hole is not a singularity that can be reached
by physical signals in a finite amount of proper time. In this sense it is
{\it not really a singularity}. } }

{\ {\ If the particle starts initially in the black hole region $u_0v_0<1,\,$
either inside or outside the horizon$,$ its trajectory has wildly different
behavior depending on its velocity. There are two critical ratios of the
velocities at which $\gamma =0.$ } }

\begin{itemize}
\begin{description}
\item  {\ {\ (i) If the velocities lie in the range
\begin{equation}
\begin{array}{l}
{{\left( \frac{1-\sqrt{1-u_0v_0}}{u_0}\right) ^2<\frac{\dot v_0}{\dot u_0}%
<\left( \frac{1+\sqrt{1-u_0v_0}}{u_0}\right) ^2.}} \\
\end{array}
\label{critical}
\end{equation}
then $\gamma $ is imaginary and (\ref{hit}) vanishes periodically. The
massive particle goes through the horizon and hits the }}black hole{{\
singularity at a {\it finite} value of $\tau .$ There it moves smoothly to a
second sheet of the $(u,v)$ space-time, but still with $uv<1$. It continues
its journey toward the second branch of the singularity and hits it, moving
on to a third sheet of space-time (or back to the first sheet, according to
interpretation). The journey continues endlessly from singularity to
singularity, always moving smoothly to another sheet, and always remaining
in the region $uv<1$. This behavior is similar to the behavior of geodesics
in the many worlds of the Reissner-Nordtrom black hole\footnote{{\ {\ In the
present case the worlds are pasted to each other just at $uv=1$ along the
singularity. When the metric is modified by quantum corrections \cite{ibsf}
a gap develops so that the singularity becomes unreachable while the
geodesics move from one world to the next.}}}. } }

{\ {\ (ii) If the velocities lie in the range
\begin{equation}
\frac{\dot v_0}{\dot u_0}>\left( \frac{1+\sqrt{1-u_0v_0}}{u_0}\right) ^2
\label{criticalii}
\end{equation}
then $\gamma $ is real, the motion is hyperbolic, and (\ref{hit}) vanishes
only once. Therefore, the particle hits the black hole at a finite $\tau $
only once, and moves to a second sheet where it remains for the rest of
time. } }

\item  {\ {\ (iii) If the velocities lie in the range
\begin{equation}
\frac{\dot v_0}{\dot u_0}<\left( \frac{1-\sqrt{1-u_0v_0}}{u_0}\right) ^2
\label{criticaliii}
\end{equation}
then $\gamma $ is real, the motion is hyperbolic, but (\ref{hit}) never
vanishes. Therefore, the particle never hits the black hole. } }
\end{description}
\end{itemize}


{\ {\ The string geodesics given below follow, on the average, the behavior
of the massive particle geodesics above. But, because of the oscillatory
motion we find new phenomena in the vicinity of the black hole. When the
string approaches the black hole from the $uv<1$ region, and hits the
singularity, it behaves differently than the particle: it fully penetrates
the wall to the $uv>1$ region, but then it snaps back into the $uv<1$
region, and then follows more or less the particle trajectory in the second
sheet, etc. (see the solution below and the plots in Figs.4,5). } }

{\ {\ To construct the string solution we use the general formulas of the
previous sections. The functions $\bar F,\bar W$ corresponding to the flat
space-time metric $G=(1-uv)^{-1}$ are given in (\ref{FWflat}). Using them in
the general formulas (\ref{boundary}-\ref{init}) we obtain the explicit
recursion relations
\begin{equation}
\begin{array}{c}
\bar W_k(\sigma ^{+},\sigma ^{-})=\frac 1{v_k}\left[ 1-\frac{\left(
1-U_k(\sigma ^{+})v_k\right) \ \left( 1-U_k(\sigma ^{-})v_k\right) }{%
1-u_{k-1}v_k}\right]  \\
U_{k+1}(z)=\frac{1-u_kv_k}{1-u_{k-1}v_k}\left[ U_k(z)+\frac{u_k-u_{k-1}}{%
1-u_kv_k}\right]  \\
u_{k+1}=\frac{2u_k-u_{k-1}-u_k^2v_k}{1-u_{k-1}v_k},
\end{array}
\label{recsl}
\end{equation}
and similarly $W_k,V_k,v_k$ are obtained from the above by interchanging $%
U\leftrightarrow V$ and $u\leftrightarrow v.$ This agrees with the results
of \cite{ibjs}. Note that for $u,v\rightarrow 0$ or $\infty $ the metric
approaches the flat metric (see footnote \#2). In both of these limits the
formulas in (\ref{recsl}) approach the flat ones in (\ref{recflat}). } }


{\ {\ Just as the flat case, we define a lattice version of the area element
in curved space-time. The ``lattice area'' swept by the string for one of
the rectangles in Fig.4,5 is defined as
\begin{equation}
dA_k=\frac{(u_k-u_{k-1})\,(v_k-v_{k-1})}{1-\frac 14(u_k+u_{k-1})%
\,(v_k+v_{k-1})}.  \label{minimsl}
\end{equation}
As in the flat case, this is a lattice version of the target space area of
the image of a $A$ or $B$ cell on the world sheet, while the area of the
image of a $C$ or $D$ cell is zero. This expression is invariant under the
``transfer matrix'' (\ref{recsl}), i.e. $dA_k=dA_{k+1}$. The invariance of
this expression everywhere, including in the vicinity of the black hole
singularity, is helpful in understanding the reason for the tunnelling to
the bare singularity region. Namely, since the string must move in a way
that conserves this minimal area, and must have a continuous trajectory, it
cannot avoid the tunnelling for generic initial conditions set by an
observer (see Fig.5). } }

{\ {\ By feeding the recursion relations to a computer, the trajectories of
the folds are plotted in Fig.4,5. A physical discussion of the string
falling into a black hole was given in \cite{ibjs}. The most surprising
effect was the tunnelling of the string into the bare singularity region
which is not possible for particles (Fig.5). As suggested before, this is
analogous to the diffraction of classical light waves that is possible for
waves but not for particles. } }

Therefore, we must conclude that a complete black hole {\it classical}
spacetime must include the ``bare singularity'' region which is a region
dual to the exterior of the horizon. We have shown that the classical motion
of a string is incomplete without this region.. Of course, we are also aware
that the quantum conformal field theory also has duality properties that
require the inclusion of all dual regions on an equal footing. All this
seems to indicate that the problems surrounding gravitational singularities,
including the information paradox problem, probably could not be fully
resolved or even correctly described without the inclusion of spacetime
regions that seemed inaccessible in traditional approaches.

\section{Cosmological space-time}

{\ {\ Consider the cosmological space-time corresponding to a Friedman -
Robertson - Walker (FRW) universe in 4D
\begin{equation}
ds^2=dt^2-R^2(t)\,\left( \frac{dr^2}{1-kr^2}+r^2d\Omega ^2\right) .
\label{FRW}
\end{equation}
where $k=-1,0,1$ are related to the classification of cosmological
space-times as ``open, flat, closed'' respectively. For a string moving
purely along the radial direction $d\theta =d\phi =0$ one concentrates on
the 2D metric
\begin{equation}
ds^2=dt^2-R^2(t)\frac{dr^2}{1-kr^2}.  \label{frw2d}
\end{equation}
It is convenient to change variables
\begin{equation}
\begin{array}{c}
\sqrt{k}\,r=\sin (\sqrt{k}X),\quad T=\int^t\frac{dt^{\prime }}{R(t^{\prime })%
},\quad \sqrt{k}=i,0,1 \\
u=\frac 1{\sqrt{2}}(T+X),\quad v=\frac 1{\sqrt{2}}(T-X),
\end{array}
\label{rX}
\end{equation}
so that the line element takes the conformal form
\begin{equation}
\begin{array}{c}
ds^2=R^2\left( dT^2-dX^2\right) =G(u,v)\,dudv, \\
G(u,v)=2R^2(t(T)).
\end{array}
\label{cosmoconf}
\end{equation}
Once written in terms of $(u,v)$ the complete manifold is usually obtained
by analytic continuation to all values of these variables. Then one may
apply the general formulas of the previous sections to obtain the classical
motion of strings. } }

{\ {\ As an example consider the de Sitter universe for which the expansion
factor of the universe is given by
\begin{equation}
|R(t)|=e^{Ht}  \label{expan}
\end{equation}
where $H=\dot R/R$ is the Hubble constant, and
\begin{equation}
ds^2=\frac{4\,du\,dv}{H^2(u+v)^2}.  \label{desit}
\end{equation}
This 2D space can be embedded in 3D as the surface of a hyperboloid
described by (see Fig.6)
\begin{equation}
\begin{array}{c}
x_0^2-x_1^2-x_2^2=-H^{-2} \\
x_0=\frac{uv-H^{-2}}{u+v},\quad x_1=\frac{uv+H^{-2}}{u+v},\quad x_2=\frac 1H%
\frac{u-v}{u+v}
\end{array}
\label{D23D}
\end{equation}
Then the metric in (\ref{desit}) takes the flat form
\begin{equation}
ds^2=dx_0^2-dx_1^2-dx_2^2.  \label{flatdesit}
\end{equation}
} }


{\ {\ First consider the geodesic equations for a massive particle of mass $m
$ . They can be solved exactly as a function of proper time $\tau $%
\begin{equation}
\begin{array}{c}
u(\tau )=c+\frac{\sinh (Hm\tau )-\sinh (Hm\tau _0)}{H\sinh [Hm(\tau +\tau
_0)]} \\
v(\tau )=-c-\frac{\sinh (Hm\tau )+\sinh (Hm\tau _0)}{H\sinh [Hm(\tau +\tau
_0)]} \\
R(\tau )=\frac{\sinh [Hm(\tau +\tau _0)]}{\sinh (Hm\tau _0)}=-\frac{\sqrt{2}}%
H\frac 1{u+v}
\end{array}
\label{soldesit}
\end{equation}
where $c,\tau _0$ are constants, and $R(\tau =0)=1$ has been chosen for
simplicity. The geodesic for the massive particle is best pictured on the
surface of the hyperboloid $x_1^2+x_2^2=x_0^2+H^{-2}$. Inserting the
solution in (\ref{D23D}) one sees that $x_0(\tau )$ increases monotonically
and lies in the range $-\infty <x_0(\tau )<\infty $ . The geodesic extends
from a point on the infinitely large circle at $x_0=-\infty $ to a point on
the infinitely large circle at $x_0=\infty .$ It is a line that spirals less
than or equal to one time on this surface. Define the angle $\tan \theta
=x_2/x_1.$ If the mass is zero, the maximum spiralling angle $\Delta \theta
=\theta (\infty )-\theta (-\infty )$ is exactly $2\pi ,$ but for the massive
particle the angle is less than $2\pi .$ } }

{\ {\ Thus, on the average, we must expect the string center of mass to
spiral less than $2\pi .$ Of course, the overall string performs the yo-yo
oscillations of Fig.3 and sweeps a minimal surface on the hyperboloid, that
is similar to the one in flat space-time except for deformations due to
curvature. } }

{\ {\ The explicit solution that describes this motion is obtained by
applying our general procedure that yields the transfer matrix
\begin{equation}
\begin{array}{c}
\bar W_k=\left[ \frac 1{U_k(\sigma ^{+})+v_k}+\frac 1{U_k(\sigma ^{-})+v_k}-%
\frac 1{u_{k-1}+v_k}\right] ^{-1}-v_k \\
U_{k+1}(z)=\left[ \frac 1{U_k(z)+v_k}+\frac 1{u_k+v_k}-\frac 1{u_{k-1}+v_k}%
\right] ^{-1}-v_k \\
u_{k+1}=\left[ \frac 2{u_k+v_k}-\frac 1{u_k+v_{k-1}}\right] ^{-1}-v_k
\end{array}
\label{transdesit}
\end{equation}
Similar formulas hold for $W_k,V_k,v_k$ respectively. We define a discrete
version of the minimal area for rectangle $k$ by
\begin{equation}
dA_k=\frac 4{H^2}\frac{(u_k-u_{k-1})\,(v_k-v_{k-1})}{%
(u_k+v_{k-1})(u_{k-1}+v_k)}.  \label{minimdesit}
\end{equation}
The transfer matrix leaves invariant this discrete minimal area, i.e. $%
dA_{k+1}=dA_k$. This is easily proven by rewriting the transfer matrix in
the form
\begin{equation}
\begin{array}{c}
\frac{U_{k+1}(z)-u_k}{[U_{k+1}(z)+v_k](u_k+v_k)}=\frac{U_k(z)-u_{k-1}}{%
[U_k(z)+v_k](u_{k-1}+v_k)} \\
\\
\frac{V_{k+1}(z)-v_k}{[u_k+V_{k+1}(z)](u_k+v_k)}=\frac{V_k(z)-v_{k-1}}{%
[u_k+V_k(z)](u_k+v_{k-1})}.
\end{array}
\label{transdesitt}
\end{equation}
By feeding the recursion relation to a computer, and plotting the
trajectories of the folds, the minimal surface is constructed and seen to
have the properties described above, as depicted in Fig.6.}}

\section{Quantum Folded String}

{\ Given the fact that the string in 2D is quite non-trivial classically, we
expect that there is a consistent quantization procedure that includes the
non-trivial folded states. Therefore we should try to make a case for folded
strings in the quantum theory.}

As pointed out many times in our past work, folded 2D-string states are
present in the $d=2$ and $c\leq 25$ sector of the quantum theory in flat as
well as curved spacetime. In simple string models, when it has been possible
to compute the spectrum, their norm is positive and is proportional to $%
(c-26)$. Only if $d=2$ and $c=26$ simultaneously (e.g. $d=2$ flat space-time
with linear dilaton such that $c=26$) the folded string states become zero
norm states and then the special discrete momentum states survive as the
only stringy states. A simple model in which these properties may be easily
seen is the {\it covariant} quantization of the 2D string theory, in which
the physical states are identified as the subset that satisfies the Virasoro
constraints, i.e. $L_0-\frac{d-2}{24}=L_{n_{\geq 1}}=0$ applied on states.
For example, it has been known for a long time that the $d\leq 25$ sector of
the flat string theory has non-trivial positive norm states (including for $%
d=2)$ that satisfy the Virasoro constraints and that there are no ghosts
\cite{noghost}. {A similar covariant quantization can be carried out for the
2D black hole string by using the Kac-Moody current algebra formulation, and
relaxing the $c=26$ condition (i.e. $k<9/4$) to include the folded strings. }

Why $c=26?$ There are several approaches to the quantization of strings that
converge on the requirement of $c=26.$ These include the light-cone gauge,
the Polyakov path integral and the BRST quantization. However, they each
involve certain steps that seem to inadverdently exclude the $c<26$ string.
We can point out that

\begin{description}
\item  (i) The usual light-cone approach throws away the folded states from
the beginning by assuming a uniform momentum density $P^{+}(\tau ,\sigma
)=p^{+}$, a statement that is not true for the BBHP solutions even in flat
spacetime.

\item  (ii) The Polyakov approach assumes a certain measure for the path
integral, thus locking into a {\it definition} of a quantum theory. A
different measure that takes into account folds can be considered as in {{%
\cite{yankielowicz} mentioned below.}}

\item  {(iii) The BRST approach requires }${Q}_{BRST}^2=0$ as an operator.
This is a stronger requirement than imposing the Virasoro constraints only
in the physical subspace $<phys|L_n-\alpha _0\delta _{n0}|phys>=0$. An
analogous statement would be $<phys|Q_{BRST}|phys>=0,$ which does not lead
to $c=26.$ Actually, the fact that there exists a consistent covariant
quantization of the {\it flat free string} in $d<26$ is already proof that
the $Q_{BRST}^2=0$ approach is too strong.
\end{description}

\noindent The critical  $c=26$ string is certainly consistent and well
understood. The success of its methods have created a prejudice against
other possibilities. However, as argued above, it appears that a more
general quantization of string theory for $c<26$, that would include
classical features such as the folded string states, is possible. This is
already known to be true for the free flat string. {What would also be
interesting is to find the correct formulation for interacting folded
strings. The path integral approach }discussed{{\ in \cite{yankielowicz}
seems to be promising, and it may be possible to make faster progress by
reformulating it in the conformal gauge and relating it to our classical
solutions. \ Note that the definition of fold in ref.\cite{yankielowicz}
does not take into account that the map from the world sheet to spacetime
may be many to one\footnote{%
By definition, at a fold the determinant of the induced metric, $g_{\alpha
\beta }=\partial _\alpha x^\mu \partial _\beta x^\nu G_{\mu \nu },$
vanishes, $\det g=0$. But in the conformal gauge the induced metric itself
also vanishes locally everywhere since $g_{\alpha \beta }=\Lambda \eta
_{\alpha \beta }$. Note that this does not mean that the world sheet metric $%
\eta _{\alpha \beta }$ vanishes, rather $\Lambda =0,$ indicating that the
map is singular precisely at the fold. In the $C,D$ cells by virtue of
having either $u$ or $v$ constant throughout the cell one gets $g_{\alpha
\beta }=0,$ indicating that all points in these cells are mapped to the
trajectory of the fold in target spacetime. The mapping is many to one,
since a region of the world sheet is mapped to a segment (trajectory of the
fold) in target spacetime. Therefore, a fold in target spacetime has many
representatives on the world sheet. For example, consider the leftmost $C$%
-type cell at the bottom of the diagram in (\ref{soll}) for which $%
u=u_{k-1}, $ and $v=W_{k-1}(\sigma ^{+},\sigma ^{-})$. At a constant $\tau
=\tau _0,$ all the $\sigma $ points that give the same value of $v=v_0$ are
mapped to the same fold located at $(u_{k-1},v_0)$. As $\tau $ changes $%
u=u_{k-1}$ remains fixed while $v$ changes along the lightlike trajectory of
the fold. To trace the trajectory of a fold it is sufficient to concentrate
on one of its images on the world sheet. Such representative images are the
vertical lines at $\sigma =0,2$ in the diagram in (\ref{soll}).} (i.e. a
region maped to a segment, as is the case for our solutions)}}. {This
feature may be important in the formulation of folds and their interactions
in the path integral approach. In particular, the description of folds in
the conformal gauge, as in our papers, may eventually prove to be a more
convenient mathematical formulation.}

\section{Higher dimensions}

{\ {\ Folded strings exist in higher dimensions as well. One can display the
general solution in flat space-time in the temporal gauge
\begin{equation}
\begin{array}{c}
x^0=p^0\tau ,\,\,{\bf x}(\tau ,\sigma )={\bf x}_L(\sigma ^{+})+{\bf x}%
_R(\sigma ^{-}),\, \\
(\partial _{+}{\bf x}_L)^2=p_0^2=(\partial _{-}{\bf x}_R)^2 \\
\\
\partial _{+}{\bf x}_L=p^0\left( \frac{2{\bf f}}{1+{\bf f}^2},\frac{1-{\bf f}%
^2}{1+{\bf f}^2}\varepsilon _L\right) , \\
\partial _{-}{\bf x}_R=p^0\left( \frac{2{\bf g}}{1+{\bf g}^2},\frac{1-{\bf g}%
^2}{1+{\bf g}^2}\varepsilon _R\right)
\end{array}
\label{dfolded}
\end{equation}
where ${\bf f}(\sigma ^{+}){\bf ,\,\,g(\sigma ^{-})}$ are arbitrary periodic
vectors in $d-2$ dimensions, {\it which could be discontinuous}, and $%
\varepsilon _L(\sigma ^{+}),\,\varepsilon _R(\sigma ^{-})$ take the values $%
\pm 1$ in patches of the corresponding variables such that the sign patterns
repeat periodically (as in the 2D\ string). When ${\bf f,g}$ are both zero
the solution reduces to the 2 dimensional BBHP case. In general, the
presence of discontinuous $\varepsilon _{L,}\varepsilon _R,$ and the
discontinuities in ${\bf f}(\sigma ^{+}){\bf ,\,\,g(\sigma ^{-})}$ gives a
larger set of solutions, which include strings that are partially or fully
folded. Discontinuities are allowed since the differential equations are
first order in the derivatives }$\partial _{+}$ and $\partial _{-}.${\ }Such
solutions are usually missed in the lightcone gauge even in the flat
classical theory (therefore, the lightcone ``gauge'' is not really a gauge,
except for the case of }$c=26${). }

{\ The curved space-time analogs of such solutions in higher dimensions are
presently under investigation. For other solutions of classical strings
theory in higher dimensional curved spacetime see \cite{devega}.}

\section{Comments and conclusions{\ }}

{\ {\ We have solved generally the classical 2D string theory in any curved
space-time. All stringy solutions correspond to folded strings. All
solutions tend to the BBHP\ solutions (as boundary conditions) in the
asymptotically flat region of the curved space-time. Therefore, the BBHP
solutions of eq.(\ref{general}) serve to classify all the solutions for any
curved space-time. In fact, the sign patterns of the BBHP solutions provide
the method for dividing the world-sheet into patches, thus defining the
lattices associated with the $A,B,C,D$ solutions. The matching of boundaries
for these functions gives the general solution in curved space-time in the
form of a ``transfer matrix''. Thus, }}lattices on the world-sheet plus
geometry in space-time lead to transfer matrices. This seems to be a rich
area to explore in more detail.

{\ {\ The general physical motion of the string is: oscillations around a
center of masss that follows on the average a geodesic of a massive
particle, consistent with intuition. The oscillations are deformed by
curvature as compared to the BBHP solutions in flat spacetime, but they
maintain the same general character as long as the curvature is smooth.
However, new stringy behavior becomes evident in the vicinity of
singularities where new phenomena, such as tunneling (similar to
diffraction), take place. } There is also the continuation of the motion
into new worlds, in a finite amount of proper time, that the string as well
as the massive particle geodesics do (but not the massless particle! - see
above). Because of the tunelling and the new worlds, the global space of the
}$SL(2,R)/R$ black hole is not just the usual black hole space, $uv<1.$
Rather, it must include also the $uv>1$ ``bare singularity'' region even for
the classical description of strings (actually this region is not really
singular, as argued above). We conjecture that the inclusion of the bare
singularity region is a more general requirement than the $SL(2,R)/R$ case
for the correct description of string motion. Of course, by duality, the
quantum theory must include all the regions.

Folded strings are also of interest in a string-QCD relation. Gluons are
expected to behave just like the folds, since only at the location of a
gluon the color flux tube can fold, and move at the speed of light. Some
recent discussion on this point can be found in \cite{ibberkeley}\cite
{ibddcl}.

{We suspect that the inclusion of the quantum states corresponding to folded
strings may lead to a consistent quantum theory in less than 26 dimensions.
As already emphasized earlier in the paper, the free string is perfectly
consistent as a quantum theory for $c<26$, including the folded states. The
interacting quantum string with folds remains as an open possibility.}

\section{Figure captions}

\begin{figure}
\caption{The 2-fold solution in flat spacetime.}
\end{figure}

\begin{figure}
\caption{The 3-fold solution obtained with different left-right periods.}
\end{figure}

\begin{figure}
\caption{The two fold solution in curved spacetime.}
\end{figure}

\begin{figure}
\caption{String falling into black hole.}
\end{figure}

\begin{figure}
\caption{Bare singularity region is needed in classical string theory.}
\end{figure}

\begin{figure}
\caption{String motion in de Sitter spacetime.}
\end{figure}

\newpage
\widetext
\begin{figure}
\epsfbox{ 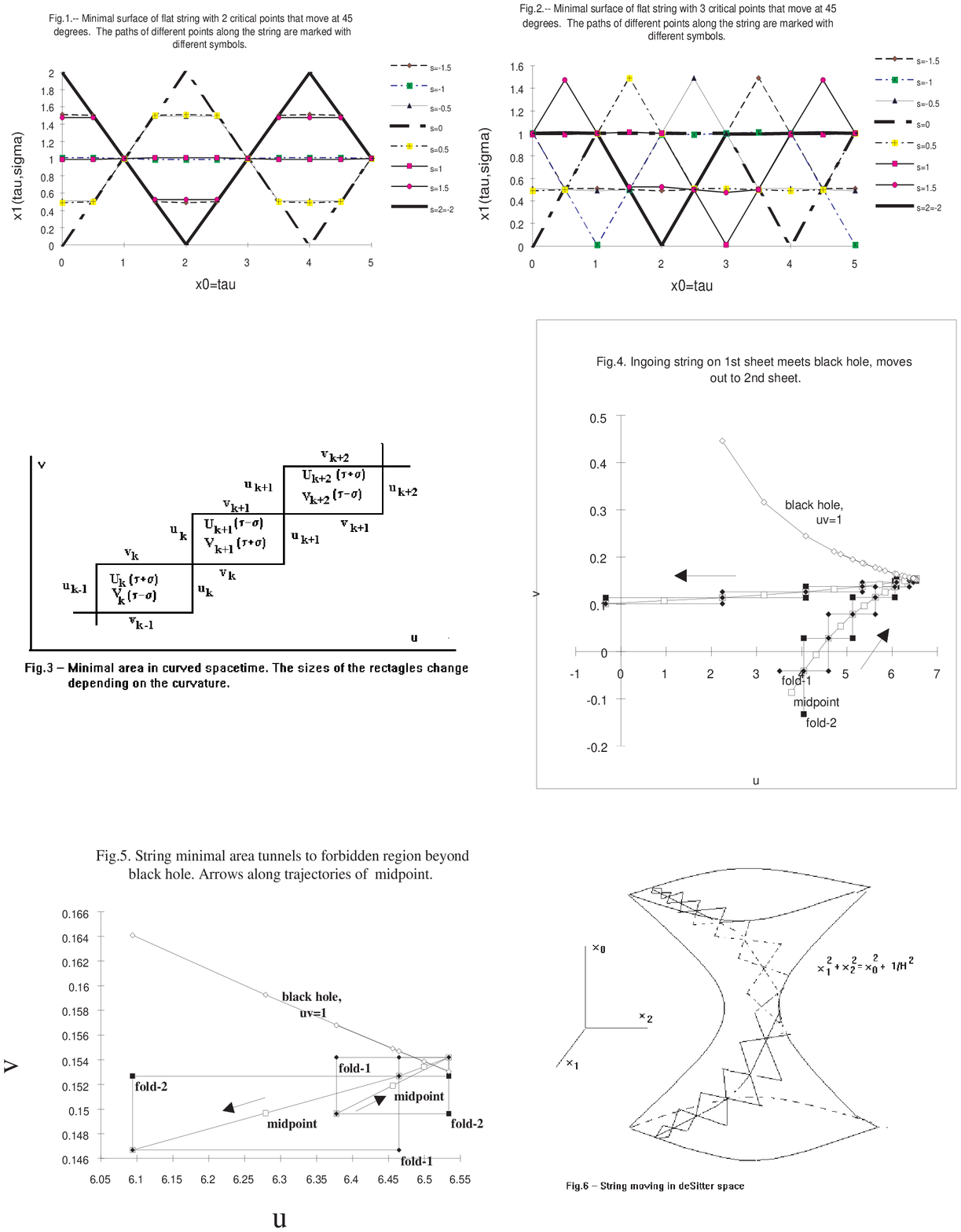}
\end{figure}

\end{document}